# Some consequences of Sica's approach to Bell's inequalities.

Alejandro A. Hnilo.

CEILAP, Centro de Investigaciones en Láseres y Aplicaciones, (CITEDEF-CONICET).
CITEDEF, J.B. de La Salle 4397, (1603) Villa Martelli, Argentina.
*email*: ahnilo@citedef.gob.ar

*Abstract*

Louis Sica derived Bell's inequalities from the hypothesis that the time series of outcomes observed in one station does not change if the setting in the other (distant) station is changed. This derivation is based on arithmetical properties only. It does not involve the controversial definitions of Locality and Realism, it does not require the definition of probabilities, and is valid for series of any length. In this paper, Sica's approach is extended to series with non ideal efficiency and to the actual time structure of experimental data. The first extension leads to an interesting relationship, involving the entanglement parameter $S_{CHSH}$ and efficiency, that places the so-called "detection loophole" under new light. The second extension makes visible that measuring with different settings unavoidably means recording series at different times. It replaces "Local Realism" (as the assumption necessary for the validity of Bell's inequalities), with the assumption that the recorded series can be arbitrarily reordered. Violation of this latter assumption is, in my opinion, more acceptable to intuition than violation of Local Realism. The second extension also shows that the observation of a violation of Bell's inequalities implies that Sica's hypothesis is not valid, i.e., that the series in one station is different if the setting in the other station is changed. This result gives precise meaning to "quantum non-locality", and also explains why it cannot be used for sending messages. Finally, it is demonstrated that a series of outcomes, even if it violates Bell's inequalities, can be always embedded in a set of factual and counter-factual data in which Sica's hypothesis is valid. In consequence, factual universe may be quantum (non-classical) or not, but the union of factual and counter-factual universes is always classical.

July 22nd, 2024.

## 1. Introduction.

Bell's inequalities have been derived in many different ways. The hypotheses involved in their derivation are also different. The best known ones can be gathered in two big groups: "Locality" and "Realism". Discussions on the precise definition, meaning, validity and consequences of these hypotheses have been the subject of myriads of papers for years. Nevertheless, there is a derivation of Bell's inequalities that is free from philosophical and mathematical ambiguities, from the definition of classical probabilities and from statistical limitations. This derivation is due to Louis Sica [1], and reviewed in the next Section 2. In this paper, some new consequences of Sica's approach are derived. They are: an interesting bound to the product of entanglement and efficiency (Section 3), an assumption which rejection is more acceptable to intuition than the rejection of Local Realism, and a clearer interpretation of "non-locality" is obtained (Section 4). Finally (Section 5) it is demonstrated that all measured data, even if they violate Bell's inequalities, can be embedded into a classical set of instructions.

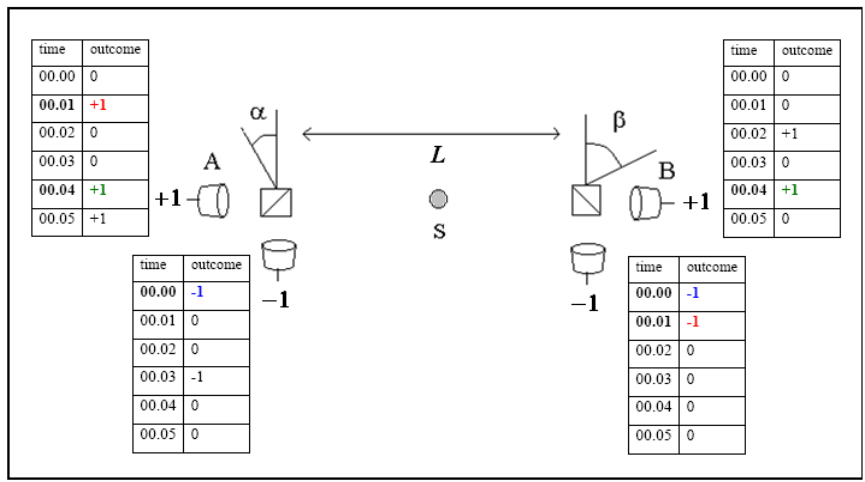

Figure 1: Sketch of a time-stamped Bell's experiment. The source S emits entangled photons that propagate towards stations A and B, which are separated by a (large) distance $L$. The "+1" (-1) in the tables means that detector "+1" (-1) fired during that time slot, "0" means that no detector fired. In this illustration, a coincidence (-,-) occurs at t=00.00, a (+,-) at t=00.01, a (+,+) at t=00.04. Single detections occur at t=00.02 (0,+), t=00.03 (-,0), etc. The rate coincidences/singles defines the efficiency $\eta$ of the corresponding detector. F.ex., the time series of detector A+ has 3 singles and 2 coincidences, then $\eta^+_A = 2/3$.

## 2. Review of Sica's derivation of Bell's inequalities ($\eta$=1).

Let consider the setup in Figure 1 with ideal efficiency ($\eta = 1$) in all detectors. This means that all detections in one of the stations are time-coincident with one detection in one of the detectors in the other station. The flux of incident particles is adjusted low (or the time slots sufficiently short), so that simultaneous detections in two detectors in the same station are zero. Let call $a_i$ ($a'_i$) the time series of outcomes recorded in station A when the angle setting is $\alpha$ ($\alpha'$), and the same for the time series $b_i$ ($b'_i$) recorded in station B when the angle setting is β (β'). Let introduce the following assumption, that I call here *Sica's condition*:

- The time series $a_i$, $a'_i$ ($b_i$, $b'_i$) are the same regardless B = β or β' (A = α or α').

which is at first sight intuitive, but involves a strong counter-factual assumption (see later). Assuming Sica's condition valid, it is possible to write down a table of series of outcomes as it is illustrated in Figure 2.

| *Time (i)* | 1 | 2 | 3 | 4 | 5 | 6 | 7 | 8 | 9 | 10 | 11 | 12 | 13 | 14 | 15 | 16 |
|---|---|---|---|---|---|---|---|---|---|---|---|---|---|---|---|---|
| $a_i$ | + | + | + | + | + | + | + | + | - | - | - | - | - | - | - | - |
| $b_i$ | - | - | + | + | + | + | + | + | + | + | - | - | - | - | - | - |
| $a'_i$ | - | - | - | - | + | + | + | + | + | + | + | + | - | - | - | - |
| $b'_i$ | - | - | - | - | - | - | + | + | + | + | + | + | + | - | - |

Figure 2: Table of a possible (although improbable) time series of *N*=16 outcomes recorded with ideal efficiency (η = 1), all single detections are also coincidences. The series $a_i$ ($a'_i$) is made of the outcomes observed at station A when the angle setting is α (α'). The series $b_i$ ($b'_i$) is made of the outcomes observed at station B when the angle setting is β (β'). As the table is fixed and fully determined, Sica's condition is valid. Note that $S_{CHSH}$ = 2; $S_{CHSH}$ > 2 is impossible in a table with η = 1 (see the text).

In the general case:

$$\left| \Sigma\, a_i.b_i - \Sigma\, a_i.b'_i \right| = \left| \Sigma\, a_i.(b_i - b'_i) \right| \leq \Sigma\, |a_i|.|b_i - b'_i| = \Sigma\, |b_i - b'_i| \qquad (1)$$

where the sum goes from *i*=1 to *N*. Here, *N* is the number of time slots, the number of single detections and also the number of coincidences in the experimental run (*N* = 16 in Fig.2). Be aware that $a_i$ = +1 or -1 $\forall i$. In the same way:

$$\left| \Sigma\, a'_i.b_i + \Sigma\, a'_i.b'_i \right| \leq \Sigma\, |b_i + b'_i| \qquad (2)$$

summing up eqs.1 and 2:

$$\left| \Sigma\, a_i.b_i - \Sigma\, a_i.b'_i \right| + \left| \Sigma\, a'_i.b_i + \Sigma\, a'_i.b'_i \right| \leq \Sigma\, |b_i - b'_i| + \Sigma\, |b_i + b'_i| \qquad (3)$$

For a given value of *i*, the first term in the rhs is 2 (0) if $b_i$ and $b'_i$ have the same (different) sign. The opposite occurs for the second term in the rhs, so that the rhs is 2 for all values of *i*. Hence:

$$(1/N).\left| \Sigma\, a_i.b_i - \Sigma\, a_i.b'_i \right| + (1/N).\left| \Sigma\, a'_i.b_i + \Sigma\, a'_i.b'_i \right| \equiv S_{CHSH} \leq 2 \qquad (4)$$

which is the Clauser-Horne-Shimony and Holt inequality (CHSH). Compare with the usual derivation, which involves averages over a space of hidden variables [2]. F.ex., in Fig.2, E(α,β) = (-2+6-2+6)/16 = ½ , the same for the other settings, and $S_{CHSH}$ = 2.

Note that Sica's derivation only involves observed (i.e., unquestionably real) time series of outcomes, and arithmetical properties. It is valid for series of any length. It is free of the discussions involving the definitions of "Locality" and "Realism", of the size of statistics or of the hypothesis of hidden variables. It does not require using probabilities either. Recall that defining classical probabilities presupposes a Boolean algebra, what is a logical inconsistency when dealing with the quantum realm [3].

On the other hand, Sica's condition is more restrictive than the usual meaning of "Locality", which only means statistical independence: $P_{A=\alpha,B=\beta}(\alpha,\beta,\lambda) = P_{A=\alpha}(\alpha,\lambda).P_{B=\beta}(\beta,\lambda)$, i.e., an equality involving *averaged* frequencies. Sica's condition, instead, requires the identity of time series term by term, or else, that the series can be arbitrarily reordered (this issue is considered in Section 4 here). It may be depicted as the strongest form of *non-contextuality* [2]. Besides, considering the series recorded in A when B=β and (simultaneously) B=β' involves a counter-factual situation, for it is impossible measuring two series with different settings at the same time (see Section 4). Nevertheless, counter-factual results can be explored in computer simulations, where the program can be run many times with identical values of the relevant variables (hidden or not). Sica's condition is then a criterion for computer programs aimed to explain "how Nature does it" [4] within the classical realm. A satisfactory classical code should be able to print sets of time series $\{a_i, a'_i, b_i, b'_i\}$ violating Bell's inequalities and holding to Sica's condition.

Before going on, let see how Sica's condition applies to the Clauser-Horne (CH) inequality. The CH setup uses only one detector per station, say the "+" ones (see Fig.1). The relevant terms in the CH inequality are:

$$J \equiv N_c(\alpha,\beta) + N_c(\alpha,\beta') + N_c(\alpha',\beta) - N_c(\alpha',\beta') - S(\alpha) - S(\beta) \tag{5}$$

where $N_c(i,j)$ is the number of "+,+" coincidences when the angle setting is $\{i,j\}$ and $S(\alpha)$ ($S(\beta)$) is the number of single detections in station A(B) with A=α (B=β). Then $a_i, b_i = \{0,1\}$ only, and $S(\alpha) = \Sigma a_i$ ($S(\beta) = \Sigma b_i$) where the sum is over the total number of time slots in the run. Using Sica's condition, then:

$$J = \Sigma (a_i.b_i + a_i.b'_i + a'_i.b_i - a'_i.b'_i - a_i - b_i) \equiv \Sigma T_i \tag{6}$$

For an arbitrary term $T_i$ in the sum: $T_i = a_i.(b_i + b'_i) + a'_i.(b_i - b'_i) - a_i - b_i$. If $a_i = 0$ and $a'_i = 1$, then $T_i = -b'_i \leq 0$ (recall $b'_i = 0$ or 1 now). If $a_i = a'_i = 0$, then $T_i = -b_i \leq 0$. If $a_i = 1$ and $a'_i = 0$, then $T_i = b'_i -1 \leq 0$. Finally, if $a'_i = a_i = 1$ then $T_i = b_i -1 \leq 0$. Therefore, $T_i \leq 0 \ \forall i \Rightarrow J \leq 0$, and the CH inequality is demonstrated from Sica's condition and arithmetical properties only. Instead, the usual derivation involves defining probabilities, the assumption of statistical independence, and proper integration over a space of hidden variables [2].

## 3. Case η < 1.

It was early recognized that a minimum value of efficiency is necessary to refute classical theories experimentally. Depending on the quantum state, the inequality and the alternative (classical) theory involved, that minimum value ranges from 2/3 [5] to 2(√2-1) ≈ 0.83 [6] or even 3-3/√2 ≈ 0.88 [7]. This necessity is the consequence of the possible existence of a conspiratorial classical mechanism generally known as the "fair sampling" or "detection" loophole. After much effort, sophisticated experiments reached the required efficiency limit and confirmed the violation of Bell's inequalities [8-13]. I have always considered the "detection loophole" a sort of artificial argument. I concurred with J.S.Bell's opinion that: *"…it is hard for me to believe that QM works so nicely for inefficient practical set-ups and is yet going to fail badly when sufficient refinements are made"* [14]. The results to be discussed in this Section make me doubtful now.

If η<1, some elements in the time series are zero. In Figure 3, just one element in each series (which are otherwise the same as in Fig.2) is zero and the zeroes do not coincide, then η = 14/15 <1 in all the series. Yet, E(α,β) = (-1+6-2+5)/14 = 8/14; the same holds for the other settings, and $S_{CHSH}$ = 32/14 ≈ 2.29 > 2, hence violating CHSH without violating Sica's condition. The key for this result is that elements in (say) series $a_i$ that do not produce coincidences when correlated with series $b_i$ (as in $i$=11) are now free to produce coincidences when correlated with series $b'_i$.

| Time (i) | 1 | 2 | 3 | 4 | 5 | 6 | 7 | 8 | 9 | 10 | 11 | 12 | 13 | 14 | 15 | 16 |
|---|---|---|---|---|---|---|---|---|---|---|---|---|---|---|---|---|
| $a_i$   | 0 | + | + | + | + | + | + | + | - | - | - | - | - | - | - | - |
| $b_i$   | - | - | + | + | + | + | + | + | + | + | 0 | - | - | - | - | - |
| $a'_i$  | - | - | - | - | 0 | + | + | + | + | + | + | + | - | - | - | - |
| $b'_i$  | - | - | - | - | - | - | + | + | + | + | + | + | + | + | 0 | - |

Figure 3: The outcome "0" means that no photon is detected in that time slot. In this illustration there is only one "0" in each series and there are no coincident "0", then η = 14/15 < 1 in all the series. The series now violate CHSH without violating Sica's condition; here $S_{CHSH}$ = 32/14 ≈ 2.29 > 2.

The question now is: what is the form of CHSH derived from Sica's condition if η<1? Consider eq.3, which now reads:

$$|\Sigma a_i.b_i - \Sigma a_i.b'_i| + |\Sigma a'_i.b_i + \Sigma a'_i.b'_i| \leq \Sigma |a_i|.|b_i - b'_i| + \Sigma |a'_i|.|b_i + b'_i| \quad (7)$$

where now the sums are over the number of coincidences $N_c$ (for, only terms with *both* factors different from zero add to the sum). For simplicity, $N_c$ is assumed the same for all angle settings (besides, this feature is desirable in experiments). The lhs is then $N_c.S_{CHSH}$. The rhs requires some analysis instead. Let define the following sets (note the bold typing) of elements of the time series:

**α** is the set made of the terms of $a_i$ that are different from 0, i.e.: **α** ≡ {$i$ / $a_i$ ≠ 0}.

**α'** is the set made of the terms of $a'_i$ that are different from 0, i.e.: **α'** ≡ {$i$ / $a'_i$ ≠ 0}.

β is the set made of the terms of $b_i$ that are different from 0, i.e.: $\beta \equiv \{i \,/\, b_i \neq 0\}$.

β' is the set made of the terms of $b'_i$ that are different from 0, i.e.: $\beta \equiv \{i \,/\, b'_i \neq 0\}$.

Besides, let call **Bs** (**Bd**) the set made of the terms where $b_i$ and $b'_i$ have the same (different) sign. Note that **Bs**∪**Bd** = β∩β'.

The first term in the rhs of eq.7 can be different from zero only for $i \in \alpha \cap (\beta \cup \beta')$. In the same way, the second term can be different from zero only for $i \in \alpha' \cap (\beta \cup \beta')$. But α∩(β∪β') = (α∩β)∪(α∩β') - α∩β∩β' and the same for the second term and α'. The number of elements (or $i$-values) in the set α∩β is, by definition, the total number of coincidences recorded when the setting is A=α and B=β, that is, $N_c(\alpha,\beta)$. The same applies to the sets α∩β', α'∩β and α'∩β'. The elements in the set α∩β∩β' are different from zero only if $b_i$ and $b'_i$ have different sign, that is, if they belong to **Bd**, in which case they sum twice (see eq.7). The same applies to the elements in the set α'∩β∩β' and the set **Bs**. Eq.7 then becomes:

$$N_c \cdot S_{CHSH} \leq N_c(\alpha,\beta) + N_c(\alpha,\beta') + N_c(\alpha',\beta) + N_c(\alpha',\beta') - 2 \cdot N_{\alpha \cap \beta \cap \beta' \cap Bs} - 2 \cdot N_{\alpha' \cap \beta \cap \beta' \cap Bd} \qquad (8)$$

where $N_{\alpha \cap \beta \cap \beta' \cap Bs}$ (or $N_{\alpha' \cap \beta \cap \beta' \cap Bd}$) is the number of elements in the set α∩β∩β'∩**Bs** (or α'∩β∩β'∩**Bd**).

The signs of the outcomes in station A and in station B are correlated depending on the angle setting, but there is no necessary correlation between the signs of outcomes recorded in the *same* station with different settings, say $b_i$ and $b'_i$. Therefore, the number of elements in the set α∩**Bs** and in α∩**Bd** can be assumed to be nearly the same in long series (yet, be warned that in Fig.3 $N_{\alpha \cap Bs}$= 6 and $N_{\alpha \cap Bd}$= 7 instead). It has been already assumed that the total number of coincidences $N_c$ is the same for all settings. Using these two assumptions, eq.8 simplifies to:

$$N_c \cdot S_{CHSH} \leq 4 \cdot N_c - N_{\alpha \cap \beta \cap \beta'} - N_{\alpha' \cap \beta \cap \beta'} \qquad (9)$$

Let define now $\mu \equiv N_{\alpha \cap \beta \cap \beta'}/N_c$ and $\mu' \equiv N_{\alpha' \cap \beta \cap \beta'}/N_c$ and assume, as a further simplification, that $\mu = \mu'$, which is valid in long series. The task now is to find a relationship between efficiency η (assumed the same for all detectors and settings) and μ. A graphical representation is helpful.

In the Figure 4, the set α is represented as a square. The number of elements in the square is the number of *single* detections in station A, regardless the sign. The gray areas are the sets α∩β and α∩β'; which are of the same size, for η is assumed the same for all detectors and it is also assumed that $\mu = \mu'$. Then $\eta = N_{\alpha \cap \beta} / N_\alpha$. It is visible that if $\eta \leq \frac{1}{2}$ the set β∩β' is empty, and eq.9 becomes the tautology $S_{CHSH} \leq 4$. If $\eta > \frac{1}{2}$ instead, β∩β'≠ **0** and:

$$\mu = (2.\eta - 1) / \eta \quad (10)$$

Replacing eq.10 into eq.9 using the definition of $\mu$ ($=\mu'$), this interesting expression is obtained:

$$S_{CHSH} \cdot \eta \leq 2 \quad (11)$$

which is valid only if $\eta > ½$, otherwise $S_{CHSH} \leq 4$ applies. Eq.11 has been derived before in the framework of specific hidden variables models [15]. Here, instead, eq.11 is derived from arithmetical relationships involving observable series of outcomes of any length, plus Sica's condition. No hidden variables or conspiratorial mechanisms are involved.

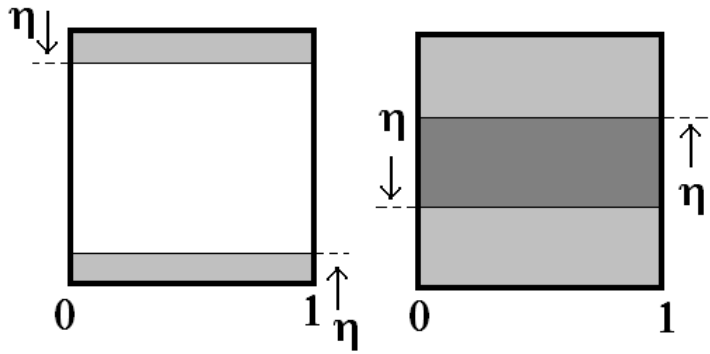

Figure 4: The squares represent the set $\alpha \equiv \{i\ /a_i \neq 0\}$. The light gray areas represent the sets $\alpha \cap \beta$ (arrow from above) and $\alpha \cap \beta'$ (arrow from below); their size (as a fraction of the area of the whole square, which is equal to 1) measure the efficiency $\eta$, which is assumed equal for all settings. Left: if $\eta \leq ½$, $\alpha \cap \beta \cap \beta' = \mathbf{0}$. Right: if $\eta > ½$ the sets $\alpha \cap \beta$ and $\alpha \cap \beta'$ intersect (dark grey). This intersection measures the rate $N_{\alpha \cap \beta \cap \beta'}/N_{singles} = (N_{\alpha \cap \beta \cap \beta'}/N_c).(N_c/N_{singles}) = \mu.\eta$, and its area is geometrically equal to 1-2.(1-$\eta$), then eq.10 follows.

Regarding the CH inequality, the elements in the series are allowed to take the value "0" from start, thus violation of CH refutes Sica's condition regardless the value of $\eta$. Yet, the value of $\eta$ does become important when considering the problem of observing that violation in practice. Assume the series recorded with $\eta=1$ do violate CH. In practice, $N_c(\alpha,\beta) = P^{++}(\alpha,\beta).N.\eta^2$ and $S(\alpha) = P^+(\alpha).N.\eta$, where $P^{++}(\alpha,\beta)$ and $P^+(\alpha)$ are the ideal probabilities of detection. This implies that a minimum value of $\eta$ must be reached in order to be able to observe the violation, even if the probabilities predicted by QM (which do violate CH) are exactly reproduced in the experiment with $\eta=1$. Depending on the quantum state and the assumed loopholes, the different bounds on $\eta$ mentioned at the beginning of this Section are obtained.

The Reader may have noted that the values in Fig.3 do not hold to eq.11: $S_{CHSH}.\eta = (32/14).(14/15) = 32/15 \approx 2.13 > 2$. This is because (as it is warned), $N_{\alpha \cap Bs} = 6 \neq N_{\alpha \cap Bd} = 7$, and one of the assumptions leading to eq.9 fails. If eq.8 is used instead, then: $14 \times S_{CHSH} \leq 4 \times 14 - 2 \times 6 - 2 \times 6 = 32$ and it follows that $S_{CHSH}.\eta < 32/14 \approx 2.29$, which does hold (2.13 < 2.29).

Note that the usual bound $S_{CHSH} \leq 2$ can be violated without violating Sica's condition as soon as $\eta < 1$. That is, in all real cases. This makes me think that the value of efficiency plays a role more fundamental than I had believed. The "detection loophole" may be not an artificial assumption after all. The condition $\eta < 1$ may be a feature real time series of outcomes *must* have in order to be able to violate CHSH. Some years ago, I concluded the performed experiments had settled the problem [16]. Now I feel that a second look to the "detection loophole" may be advisable.

**4. Measuring with different settings implies measuring at different times.**

Actually, the tables in Figs.2 and 3 are unreal. It is impossible measuring with different angle settings (say, A=$\alpha$ and A=$\alpha$') at the same time slot. A typical *distribution* of measuring times during an experimental run is shown in the upper part of Figure 5. The table of actually recorded series, as in Fig.2, is shown in the lower part. The empty boxes correspond to non-performed observations. F.ex: for time slots from $i$= 17 to 32, the setting is A=$\alpha$', so that there are *no outcomes* (neither +,-, or 0) for the series $a_i$ between $i$= 17 and $i$= 32. Note that Sica's condition is not valid in this example. F.ex., the series $a_i$ is a string of "+" when B=$\beta$' ($i$= 1 to 8), but a string of "-" when B=$\beta$ ($i$= 9 to 16).

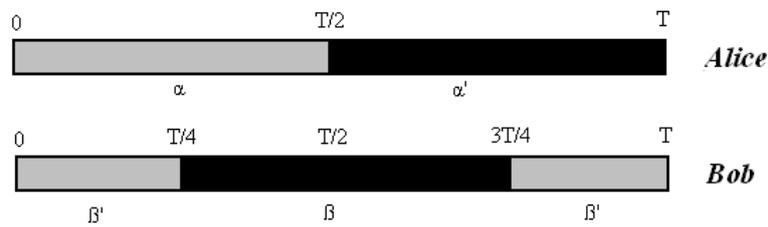

| $i$ | 1 | 2 | 3 | 4 | 5 | 6 | 7 | 8 | 9 | 10 | 11 | 12 | 13 | 14 | 15 | 16 | 17 | 18 | 19 | 20 | 21 | 22 | 23 | 24 | 25 | 26 | 27 | 28 | 29 | 30 | 31 | 32 |
|---|---|---|---|---|---|---|---|---|---|---|---|---|---|---|---|---|---|---|---|---|---|---|---|---|---|---|---|---|---|---|---|---|
| $a_i$ | + | + | + | + | + | + | + | + | - | - | - | - | - | - | - | - | | | | | | | | | | | | | | | | |
| $b_i$ | | | | | | | | | - | - | + | + | + | + | + | + | + | + | - | - | - | - | - | - | | | | | | | | |
| $a'_i$ | | | | | | | | | | | | | | | | | - | - | - | - | + | + | + | + | + | + | + | + | - | - | - | - |
| $b'_i$ | - | - | - | - | - | - | + | + | | | | | | | | | | | | | | | | | + | + | + | + | + | + | - | - |

Figure 5: Up: Typical distribution of the total measurement time T among the different angle settings. F.ex: A is set equal to $\alpha$ for time between t=0 and t=T/2 (gray area in line "Alice"), B is set equal to $\beta$ for time between t = T/4 and t = 3T/4 (black area in line "Bob"). Down: series of outcomes as in Fig.2, if measuring time is assigned as in the upper part of this Figure. In this case, the usual calculation leads to $S_{CHSH} = 0$. The empty boxes correspond to non-performed observations. They can be filled only after assuming a "possible world" [18], meaning that information external to the experiment must be added.

One may think that a table as Fig.2 can be retrieved from Fig.5 by *reordering* the series in such a way that Sica's condition is valid. Reordering means that the sub-indexes $i$ of the series (f.ex.) $a_i$ recorded between $T$/4 and $T$/2 (i.e, when B=$\beta$) are changed so that the modified series becomes equal, term by term, to the series $a_i$ recorded between 0 and $T$/4 (i.e, when B=$\beta$'). For more or less balanced series (not the case of Fig.5) this reordering seems always possible. In the worst case, a statistically irrelevant number of elements in the series should be discarded. After reordering, the table would have redundant information (for all series appear twice), its size could be halved and the empty boxes eliminated. This is called *condensing* the table. In a condensed table (say, Fig.2) Bell's inequalities hold (if $\eta<1$, Eq.11 holds). However, reordering the series is not always possible. In order to see why it is so, let consider a (balanced) factual series of length $N=4$, as in Figure 6.

| Time (i) | 1 | 2 | 3 | 4 | 5 | 6 | 7 | 8 |
|---|---|---|---|---|---|---|---|---|
| $a_i$ | - | + | - | + | | | | |
| $b_i$ | | | - | + | - | + | | |
| $a'_i$ | | | | | - | + | - | + |
| $b'_i$ | + **(-)** | - **(+)** | | | | | - | + |

Figure 6: Possible table of factual series of length $N=4$, in black (distribution of time as in the upper part of Fig.5). Sica's condition does not hold (see the series $b'_i$ at $i=1,2$ and $i=7,8$) and $S_{CHSH} = 4$. Results in brackets (and red color) can be obtained only after *arbitrary* reordering the series, and hold to Sica's condition.

Let suppose that $a_i = (-,+)$ is measured for time slots $i= 1,2$ (i.e., when B= $\beta'$), and that $b'_i = (+,-)$ is simultaneously measured, so that $E(\alpha,\beta')= -1$. In order to hold to Sica's condition, $a_i$ for $i=3,4$ (i.e., when B= $\beta$) must be also $(-,+)$. This does not necessarily occur in reality, but it can be achieved by reordering the elements of the actually measured series $a_i$. Be aware that the correlations between measured outcomes cannot change by reordering; if the series in one station is reordered, the series in the other station must be reordered in the same way. Let suppose then that in $i=3,4$ the series $b_i$ is simultaneously measured to be $(-,+)$, so that $E(\alpha,\beta)= 1$. In order to hold to Sica's condition, measured $b_i$ in $i=5,6$ must be reordered (if necessary) to be $(-,+)$ too. Let suppose that $a'_i$ is simultaneously measured to be $(-,+)$ in $i=5,6$ so that $E(\alpha',\beta)= 1$. At $i=7,8$, $a'_i$ is reordered (if necessary) to $(-,+)$ to fit Sica's condition. At this point, *if $b'_i$ is simultaneously measured to be $(-,+)$*, then $E(\alpha',\beta')=1$ and $S_{CHSH} = 4$. But, in this case, we would get $b'_i = (+,-)$ for time slots $i= 1,2$ (A= $\alpha$) and $b'_i = (-,+)$ for time slots $i= 7,8$ (A= $\alpha'$), violating Sica's condition. There is no possibility of further reordering.

If Sica's condition is enforced to hold, by *arbitrarily* imposing (f.ex.) $b'_i$ to be $(-,+)$ in time slots $i= 1,2$ (in brackets, in red, see Fig.6) without simultaneously reordering $a_i$, then $E(\alpha,\beta') = 1$ and $S_{CHSH} = 2$. But this would mean changing the measured correlation (between $\alpha$ and $\beta'$), which is something that legitimate reordering cannot do. Instead, if $a_i$ was simultaneously reordered to keep E(α,β') = -1, then all series would have to be reordered in cascade in order to hold to Sica's condition, and at the end it would be $b'_i$ = (+,-) for time slots $i$= 7,8. As $b'_i$ = (-,+) was forced for slots $i$= 1,2, Sica's condition would be violated again (and, once again, $S_{CHSH}$ = 4).

The series in Fig.6 with values in black, empty boxes (and after all reordering consistent with assumed observed correlations is made) do not hold to Sica's condition. In this sense, the black series in Fig.6 are "not classical". Yet, the difference between the series of (black) outcomes $b'_i$ when $i$ =1,2 (A= α) and when $i$ =7,8 (A= α') does not need a "spooky action at a distance" to be explained. The difference is explained simply because the series are measured at different times. There is no evident reason for Sica's condition to hold, and hence, to conclude that it should be $S_{CHSH} \leq 2$. The situation may be described by saying that *time* plays here the role of the context [2]. The same reasoning applies to the CH inequality.

In few words: in Sica's approach, the experimentally observed violation of Bell's inequalities is explained by recognizing that *arbitrary* reordering the time series is not always possible. In my opinion, this assumption is far more acceptable to intuition than giving up Locality or Realism in the usual derivation. Note that this result is valid even if η=1.

In the case the red outcomes in Fig.6 are observed instead, Sica's condition holds and Fig.6 can be condensed to Figure 7, where CHSH is obviously valid:

| Time (i) | 1 | 2 |
|---|---|---|
| $a_i$ | - | + |
| $b_i$ | - | + |
| $a'_i$ | - | + |
| $b'_i$ | - | + |

Figure 7: table obtained by condensation of Fig.6, if the outcomes that hold to Sica's condition (the ones in red) are the measured ones.

In summary: if the series of outcomes are such that Sica's condition holds (red outcomes in Fig.6), then the table can be condensed (as in Fig.7) and $S_{CHSH} \leq 2$. If, on the contrary, $S_{CHSH} > 2$ is observed, then Sica's condition does not hold and the table of outcomes cannot be condensed. Violation of Sica's condition implies that the series in one station is different if the setting in the other station is changed. This effect is in fact observed in successful numerical simulations of Bell's experiment [17]. In my opinion, the difference between the series, depending on the setting in the other station, is the best way to understand the meaning of "quantum non-locality". Nevertheless, be aware that the difference is between factual and counter-factual series, a difference that is fatally unobservable. This is, also in my opinion, the best way to understand why "quantum non-locality" cannot be used to send signals. Note that this reasoning does not involve hidden variables, probabilities, statistical independence, no-cloning theorem or any of the sophisticated (and hence, vulnerable) reasoning involved in the usual discussions on quantum non-locality and faster than light signaling. The conclusions here obtained are, in consequence, more robust and reliable than in the usual discussions.

**5. Filling the empty boxes.**

In Fig.5, the sums in Eqs.1 and 2 used to derive CHSH are disjoint. Therefore, *nothing* can be said about the value of $S_{CHSH}$ as it is usually measured [18]. No bound can be established, other than the tautology $S_{CHSH} \leq 4$. In order to put the elements of the series under the same sum, so as to derive some non trivial bound for $S_{CHSH}$, the empty boxes in Figs.5 or 6 (which correspond to counter-factual values) must be somehow filled. To assign numerical values to these empty boxes, a "possible world" must be defined to ensure logical consistency [19]. There is no mystery in this. Defining a "possible world" just means that information, additional to what is actually observed, is somehow provided. In the usual derivation (using probabilities), the Bell's inequalities are retrieved, or not, depending on the possible world chosen [18,20].

Note that, because of the results in Section 2, *any* filling of the counter-factual boxes with outcomes {+,-} drafts a table which holds to CHSH and CH. If "0" is added to the set of possible outcomes, then Eq.11 applies. If the somehow "intuitive" possible world is chosen, which assigns "0" to the counter-factual values (i.e., if the empty boxes are filled with "0"), then $\eta \leq ½$, $S_{CHSH} \leq 4$, and QM does not violate the bound. Something similar happens with the CH inequality [20]. Therefore, in order the table (with all boxes filled) to be able to violate the bound in Eq.11, the counter-factual outcomes must not belong to the set {+,-,0}, what would be certainly strange. It seems that a factual series may be non-classical (i.e. it may violate Bell's inequalities as they are usually calculated), but that it is necessarily embedded in a (factual $\cup$ counterfactual) classical table (i.e., a table that does not violate Eq.11), see also below.

Let consider now the following "possible world": in Figure 8, the factual series (in black) $b'_i$ are (+,-) when A=$\alpha$ ($i$=1,2), and (-,+) when with A=$\alpha'$ ($i$=7,8), i.e., Sica's condition is violated (and $S_{CHSH}$ >2). But it is possible to assign counter-factual values (between brackets and in blue color) $b'_{3,4}$ = (-,+) and $b'_{5,6}$ = (+,-) so that the completed series (which includes both factual and counter-factual values now) is $b'_i$ = (+,-,-,+) for both A=$\alpha$ ($i$ from 1 to 4) and A=$\alpha'$ ($i$ from 5 to 8), then holding to Sica's condition. In the same way, the other completed series in Fig.8 are $b_i$ = (-,+,-,+) ($i$ from 1 to 4 and from 5 to 8), $a_i$ = (-,+,-,+) (for both B= $\beta$, that is $i$ from 3 to 6, and B= $\beta'$, that is $i$ =

1,2 and 7,8) and $a'_i = (+,-,-,+)$. The table made of both factual and counter-factual outcomes holds to Sica's condition. This is named here a *complete Sica's table*.

| Time (i) | 1 | 2 | 3 | 4 | 5 | 6 | 7 | 8 |
|---|---|---|---|---|---|---|---|---|
| $a_i$ | - | + | - | + | (-) | (+) | (-) | (+) |
| $b_i$ | (-) | (+) | - | + | - | + | (-) | (+) |
| $a'_i$ | (+) | (-) | (+) | (-) | - | + | - | + |
| $b'_i$ | + | - | (-) | (+) | (+) | (-) | - | + |

Figure 8: Possible *complete Sica's table* for recorded series of length $N=4$. Factual values are in black and are the same as in Fig.6, they violate CHSH. Counter-factual values (between brackets, in blue) are chosen so that Sica's condition for the complete series (factual ∪ counterfactual) is valid. F.ex.: $b'_i = (+,-,-,+)$ when $A=\alpha$, $i = 1$ to 4, and also when $A=\alpha'$, $i = 5$ to 8. This table can be condensed to one of half length.

In the Appendix it is demonstrated that, given *any* set of factual series, that may violate CHSH or not (as Fig.6, black or red), it is always possible to draft a complete Sica's table (as Fig.8). The complete Sica's table plays the role of a classical hidden variable, or classical set of instructions. The violation of CHSH by the factual series can be then interpreted as the consequence of having picked up convenient observation times from some complete Sica's table (or classical set of instructions). Nevertheless, no complete Sica's table can explain the violation of CHSH for all choices of observation times (if $\eta=1$, at least). F.ex., suppose that the factual observation of the setting $A=\alpha$, $B=\beta'$ is chosen to be at time slots $i=3,4$ instead of $i=1,2$ in Fig.8. Then $a_i = (-,+)$ and $b'_i = (-,+)$ (instead of +,- in red), then $E(\alpha,\beta') = 1$ (instead of the desired value -1) and CHSH is not violated. Therefore, a complete Sica's table (explaining violation of CHSH in classical terms) can be built only *after* the factual series (that violate CHSH) are determined. A complete Sica's table provides a classical *explanation* of the violation of CHSH in an experiment, but it cannot provide a satisfactory (i.e., violating CHSH) *prediction* of outcomes for all arbitrary, still unperformed experiments.

In other words: the fact that a complete Sica's table (actually, a lot of them) always exists demonstrates that any series of outcomes violating CHSH could have been produced by a classical set of instructions. Yet, it is not possible to present a complete Sica's table where *all* arbitrarily chosen (factual) series violate CHSH. This impossibility of presenting a table of outcomes able to reproduce QM predictions for all possible choices of observations is evidently related with Kochen-Specker and GHZ theorems.

| Time (i) | 1 | 2 | 3 | 4 |
|---|---|---|---|---|
| $a_i$ | - | + | - | + |
| $b_i$ | (-) | (+) | - | + |
| $a'_i$ | (+) | (-) | - | + |
| $b'_i$ | + | - | (-) | (+) |

Figure 9: Condensation of the table in Fig.8. Note it includes both factual and counter-factual outcomes.

As said before, a table that holds to Sica's condition has redundant information and can be condensed to one of half length, eliminating the empty boxes. A complete Sica's table drafted from a factual series that violate CHSH can be therefore condensed, but the condensed table is a mixture of factual and counter-factual outcomes. See, f.ex., Figure 9. In other words: if the condensed table of the series of factual results does not include counterfactual outcomes, then the factual results do not violate CHSH.

**Conclusions.**

The usual derivation of Bell's inequalities using hidden variables and probabilities is replaced by considering the actual time series of outcomes, plus the condition that the series recorded in one station remains the same if the setting in the other station is changed (Sica's condition). This condition suffices to derive Bell's inequalities in the case of ideal efficiency. In the case of real efficiency, the interesting inequality Eq.11 is derived. In my opinion Eq.11, which arises from arithmetical properties only, shines new light on the meaning and importance of the so-called "detection loophole".

Sica's approach also makes easily visible the (often unnoticed) unavoidable limitation that measuring with different settings requires measuring at different times. This limitation implies that Bell's inequalities are valid only if arbitrary reordering of the series is assumed possible. Giving up this assumption is (in my opinion) far more acceptable to intuition than the usual alternative of giving up Locality or Realism. It means accepting that series recorded at different times are essentially different, a sort of "time contextuality".

If a given table of actually measured outcomes can be reordered so that Sica's condition holds, then it can be condensed, and Bell's inequalities are not violated. Inversely, if Bell's inequalities are observed to be violated, then the series cannot be reordered, the table of outcomes cannot be condensed, and Sica's condition does not hold, i.e., the series in A (B) are different if the setting in B (A) is changed. This result gives the words "quantum non-locality" a precise meaning. Besides, it explains why "non-locality" cannot be used for faster than light signaling: the difference is between factual and counter-factual series, a difference which is fatally unobservable (excepting in numerical simulations).

In the Appendix is demonstrated that, given any factual series of outcomes (from which a table as in Fig.6 can be built) it is possible to choose counter-factual outcomes such that the complete table holds to Sica's condition (as in Fig.8). There are $2^{N/2}$ of these tables for each set of factual series, where *N/2* is the length of the series actually recorded for each local setting in each station (f.ex. in Fig.5, *N/2* = 16). Therefore, even the series that violate CHSH can be interpreted as the consequence of having chosen appropriate observation times from a classical table (a table that holds to Sica's condition). Yet, no table can produce series that violate CHSH for *any* choosing of the observation times (if $\eta=1$, at least). Anyway, as a complete Sica's table always exists (actually, a lot of them), it is possible to conclude that the factual (observable) world may be quantum, but that the union of factual and counter-factual worlds is classical (for the Bell's experiment, at least).

As a summary:
1) Section 2: Sica's derivation of Bell's inequalities is reviewed. This derivation is based only on arithmetical properties, and the assumption that the series observed in one station does not change if the setting in the other station in changed (*Sica's condition*).
2) Section 3: If $\eta<1$, then the bound in Eq.11 applies. This result suggests that the "detection loophole" may have a role more fundamental than previously thought.
3) Section 4: "Local Realism" can be replaced (in order to derive Bell's inequalities) with the assumption that series of real data can be arbitrarily reordered. If they cannot, then Sica's condition is not valid. This is a new and (in my opinion) clearer way to understand the meaning of "quantum nonlocality", and why it cannot be used to send messages.
4) Section 5: It is always possible to add counterfactual data to the (half-empty) table of factual data, in such a way that the whole table holds to Sica's condition. Therefore, even "non-classical" observations (i.e., that violate Bell's inequalities) can be embedded into a classical description (i.e., where Bell's inequalities holds). That table can be then condensed but, if the factual data violate Bell's inequalities, then the condensed table is a mixture of factual and counter-factual data.

An interesting problem that remains to be studied is the combination of the two main features discussed in this paper, that is: the problem of an actual table of outcomes (i.e., with empty boxes) *and* $\eta<1$. This problem is the closest to the real situation. Recall that events "not observed" (counter-factual) and "zero photon observed" (which lead to $\eta<1$) are different in logical terms.

As it is just glimpsed, Sica's approach opens new and promising ways to explore the meaning and consequences of Bell's experiments.

**Appendix.**

*Proposition: A complete Sica's table can be draft from any set of factual series of outcomes.*

The demonstration is constructive.

It is evident that, as far as the observed pairs of outcomes are not changed, any distribution of the time intervals assigned to the settings can be reordered to the distribution in the upper part of Fig.5. Therefore, the discussion can deal with this distribution of time intervals. This choice has no other purpose and effect than to make notation simpler. The derivation that comes next is valid for any distribution of the time intervals assigned to the settings (f.ex: a random assignment, as in the loophole-free experiments).

For that distribution then, the set of factual (actually observed) series of outcomes is, for each angle setting in each station:

$$(a_1...a_{N/2}); (a'_{N/2}...a'_N); (b_{N/4}...b_{3N/4}); (b'_1...b'_{N/4}) \cup (b'_{3N/4}...b'_N)$$

where, for simplicity, all series are assumed of equal length ($N/2$). The counter-factual series (indicated in bold typing and blue color) are (see upper part of Fig.5):

$$(\mathbf{a_{N/2}...a_N}); (\mathbf{a'_1...a'_{N/2}}); (\mathbf{b_1...b_{N/4}}) \cup (\mathbf{b_{3N/4}...b_N}); (\mathbf{b'_{N/4}...b'_{3N/4}})$$

The complete series (factual $\cup$ counter-factual) in Alice when $A=\alpha$ and $B=\beta$ is:

$$(a_{N/4}...a_{N/2}) \cup (\mathbf{a_{N/2}...a_{3N/4}}) \tag{A1}$$

and when $B=\beta'$ is:

$$(a_1...a_{N/4}) \cup (\mathbf{a_{3N/4}...a_N}) \tag{A2}$$

In the same way, the complete series when $A=\alpha'$ and $B=\beta$ is:

$$(\mathbf{a'_{N/4}...a'_{N/2}}) \cup (a'_{N/2}...a'_{3N/4}) \tag{A3}$$

and when $B=\beta'$:

$$(\mathbf{a'_1...a'_{N/4}}) \cup (a'_{3N/4}...a'_N) \tag{A4}$$

When $B=\beta$ and $A=\alpha$:

$$(\mathbf{b_1...b_{N/4}}) \cup (b_{N/4}...b_{N/2}) \tag{A5}$$

when $B=\beta$ and $A=\alpha'$:

$$(b_{N/2}...b_{3N/4}) \cup (\mathbf{b_{3N/4}...b_N}) \tag{A6}$$

when $B=\beta'$ and $A=\alpha$:

$$(b'_1...b'_{N/4}) \cup (\mathbf{b'_{N/4}...b'_{N/2}}) \tag{A7}$$

and finally, when $B=\beta'$ and $A=\alpha'$:

$$(\mathbf{b'_{N/2}...b'_{3N/4}}) \cup (b'_{3N/4}...b'_N) \tag{A8}$$

Sica's condition requires that A1=A2, A3=A4, A5=A6 and A7=A8.

In order to make A1=A2, $(a_1...a_{N/4})$ is reordered to be equal to $(a_{N/4}...a_{N/2})$. For long nearly balanced series this should be possible discarding a statistically irrelevant number of elements. As the correlations must remain the same, the reordering forces changes in $(b'_1...b'_{N/4})$, which now becomes a new series: $(b'_1...b'_{N/4})^R$. The counter-factual (empty boxes) series $(a_{N/2}...a_{3N/4})$ and $(a_{3N/4}...a_N)$ remain free, but must be chosen to be equal between them in order to make A1=A2. F.ex.: one series is freely chosen (there are hence $2^{N/4}$ possible choices), and the other one must be equal to the freely chosen one.

In order to make A3=A4, $(a'_{N/2}...a'_{3N/4})$ is reordered to be equal to $(a'_{3N/4}...a'_N)$, and hence the series $(b_{N/2}...b_{3N/4})$ changes to $(b_{N/2}...b_{3N/4})^R$. As before, the series $(a'_{N/4}...a'_{N/2})$ and $(a'_1...a'_{N/4})$ must be equal between them but are otherwise free, so there are again $2^{N/4}$ possible choices (for arbitrarily balanced series).

In order to make A5=A6, it suffices to choose the counter-factual series $(b_1...b_{N/4})$ (the empty boxes, which are free) equal to the factual series $(b_{N/2}...b_{3N/4})^R$, and $(b_{3N/4}...b_N)$ equal to $(b_{N/4}...b_{N/2})$. In the same way, to make A7=A8, $(b'_{N/4}...b'_{N/2})$ is chosen equal to $(b'_{3N/4}...b'_N)$ and $(b'_{N/2}...b'_{3N/4})$ equal to $(b'_1...b'_{N/4})^R$ (recall it was reordered in order to make A1=A2).

A complete Sica's table is therefore built, and the proposition is demonstrated. Besides, there are $2^{N/4} \times 2^{N/4} = 2^{N/2}$ complete Sica's tables for each set of factual series, where $N/2$ is the length of the series actually recorded for each local setting in each station. F.ex in Fig.5, $N/2 = 16$.

**Acknowledgements.**

This work received support from grants N62909-18-1-2021 Office of Naval Research Global (USA), and from PIP 00484-22 and PUE 229-2018-0100018CO, both from CONICET (Argentina).

**References.**

[1] L.Sica, "Bell's inequalities I: an explanation for their experimental violation", *Opt. Commun.* **170** p.55 (1999), and "Bell's inequalities II: Logical loophole in their interpretation", *ibid* p.61.
[2] J.Clauser and A.Shimony, "Bell's theorem: experimental tests and implications", *Rep. Prog. Phys.* **41** p.1881 (1978).
[3] S.Fortín *et al.*, "A logical approach to the Quantum-Classical transition", in *Quantum Worlds: Perspectives in the ontology of Quantum Mechanics*, O.Lombardi *et al.* Editors, Cambridge University Press, Cambridge 2019.
[4] N. Gisin, "Quantum nonlocality: how does nature do it?", *Science* **326**, p.1357 (2009).
[5] P.Eberhard, "Background level and counter efficiencies required for a loophole-free Einstein-Podolsky-Rosen experiment"; *Phys.Rev.A* **47** p.R747 (1993).
[6] J.Clauser and M.Horne, "Experimental consequences of objective local theories", *Phys. Rev. D* **10** p.526 (1974).
[7] J.Larsson and R.Gill; "Bell's inequality and the coincidence-time loophole"; *Europhys.Lett.* **67** p.707 (2004).
[8] M.Giustina *et al.*, "Bell violation using entangled photons without the fair-sampling assumption", *Nature* **497** p.227 (2013).
[9] B.Christensen *et al.*, "Detection-loophole-free test of quantum nonlocality, and applications", *Phys.Rev.Lett* **111** 130406 (2013).
[10] M.Giustina *et al.*, "A significant loophole-free test of Bell's theorem with entangled photons", *Phys.Rev.Lett.* **115**, 250401 (2015).

[11] L.Shalm *et al.*, "A strong loophole-free test of Local realism", *Phys.Rev.Lett.* **115**, 250402 (2015).
[12] B.Hensen *et al.*, "Loophole-free Bell inequality violation using electron spins separated by 1.3 kilometres", *Nature* **526** p.682 (2015).
[13] W.Rosenfeld *et al.*, "Event-ready Bell-test using entangled atoms simultaneously closing detection and locality loopholes", *arXiv:1611.04604*.
[14] J.Bell, *Speakable and Unspeakable in Quantum Mechanics* (Cambridge University Press, Cambridge, 1987), p.139.
[15] M.Agüero *et al.*, "Testing a hypothetical transient deviation from Quantum Mechanics: preliminary results"; *J.Opt.Soc.Am B* **40**(4) C28-C34 (2023).
[16] A.Hnilo, "Consequences of recent loophole-free experiments on a relaxation of measurement independence"; *Phys.Rev.A* **95**, 022102 (2017).
[17] A.Hnilo, "About the description of physical reality of Bell's experiment"; *arXiv:2109.03086*.
[18] A.Hnilo, "Using measured values in Bell's inequalities entails at least one hypothesis additional to Local Realism", *Entropy* 2017, **19**, 180.
[19] B.d'Espagnat, "Nonseparability and the tentative descriptions of reality", *Phys. Rep.* **110** p.201 (1984).
[20] A.Hnilo, "Time weakens the Bell's inequalities"; *arXiv:1306.1383v2*.